# Virtualising Process Assessments to Facilitate Continual Service Improvement in IT Service Management

**Anup Shrestha**
School of Management & Enterprise
University of Southern Queensland
Toowoomba, Australia
Email: anup.shrestha@usq.edu.au

**Aileen Cater-Steel**
School of Management & Enterprise
University of Southern Queensland
Toowoomba, Australia
Email: aileen.cater-steel@usq.edu.au

**Mark Toleman**
School of Management & Enterprise
University of Southern Queensland
Toowoomba, Australia
Email: mark.toleman@usq.edu.au

## Abstract

The IT Service Management (ITSM) industry has defined processes as best practices in the widely-accepted IT Infrastructure Library (ITIL) framework. However, studies on the measurement of ITSM process improvement are scant. Our research addressed the dual problems of the lack of transparency and the need for efficiency in ITSM process assessment. Using the Design Science Research methodology, we developed a Software-mediated Process Assessment (SMPA) approach that enables assessment of ITSM processes. The SMPA approach includes process selection; an online survey to collect assessment data; measurement of process capability; and reporting of process improvement recommendations. We implemented a decision support system (DSS) to automate the SMPA approach and evaluated it at two IT service providers. The evaluations indicated that the SMPA approach supports decision-making on process improvements. The findings provided design knowledge of virtualisation in ITSM process assessment and how this may facilitate continual service improvement.

**Keywords**

Process Virtualisation Theory, IT Service Management, Process Assessment, Decision Support System, International Standards

# 1   Introduction

Business users rely upon IT services to accomplish their tasks. Organisations that receive quality IT services have a distinctive advantage in a competitive business environment. Academic research on IT service quality has concentrated on conducting gap analysis between customer expectations and perceived service quality. One of the most prominent Information Systems (IS) journals, *MIS Quarterly* featured several articles discussing the application of a service quality instrument SERVQUAL (Parasuraman et al. 1985) as an IT service quality measure (e.g. Dyke et al. 1999; Jiang et al. 2002; Kettinger and Lee 1994; Kettinger and Lee 2005; Pitt et al. 1995; Watson et al. 1998). Since the fundamental measure of the SERVQUAL model examines the gap between the customer's service expectation and perceived service delivery, it focuses on the extrinsic quality of IT services after the service is delivered.

Academic researchers have focused on non-process dimensions of IT service quality (Lepmets et al. 2014). Nevertheless, examining how IT service providers operate, in other words evaluating their intrinsic processes, is an important measure of IT service quality. It is important that service providers understand the service activities since processes impact service delivery (Walker et al. 2006). A process must be measurable in order to be controlled and improved (Praeg and Schnabel 2006). However, processes are prone to natural deterioration in the course of their evolution (Juran and Godfrey 1999). IT service management (ITSM) adopts the process approach principle of quality management (ISO 2012) in order to manage and improve activities as processes. Existing literature on ITSM has highlighted the lack of research on the topic of ITSM process measurement (Spath et al. 2011).

Process assessments determine process capability by checking compliance with a standard (Cortina 2010). In the ITSM industry, several frameworks and commercial offerings are available for ITSM process assessments such as Tudor IT Process Assessment (Barafort et al. 2009), ITIL self-assessment services (Rudd and Sansbury 2013) and PinkSCAN assessments (PinkElephant 2012). However, the ITIL books mention drawbacks to process assessments such as the lack of transparency and high costs (Lloyd 2011). High costs and time requirements have caused some researchers (Fayad and Laitinen 1997; Peldzius and Ragaisis 2013) to conclude that process assessments are wasteful. Moreover, there are heated discussions reported in the ITSM community against the use of existing ITSM process assessment approaches (England 2012; Kane 2012). The lack of transparency and high costs impede repeated process assessments which are important for continual service improvement (CSI).

The international standard for process assessment ISO/IEC 15504 suggests process assessment can be performed either as part of a process improvement activity or as part of a capability determination initiative (ISO/IEC 2005). Organisations value the process assessment as a benchmark to compare themselves with an international standard and as a yardstick in their process improvement journey (Juran and Godfrey 1999). However, formal process assessments for certification, such as class A CMMI appraisals and ISO/IEC 15504 certified process assessments, could be expensive operations with substantial costs and time commitment of several employees over several days (Lloyd 2011).

No concrete solution is presented in the academic and/or practitioner community to develop a standard measurement instrument that is accessible for IT service providers to assess their processes. An important benefit of using a measurement instrument is to be able to evaluate it in a more transparent manner with the ability to store measurement outcomes (Hubbard 2010). The ITIL framework specifies that "technology will need to be in place for monitoring and reporting" so that process improvement can occur (Lloyd 2011). Therefore, we identify the lack of transparency and the need for efficiency as the two research problems that we aim to solve by developing a new artefact for ITSM process assessment. The new artefact is called the Software-mediated Process Assessment (SMPA) approach. The SMPA approach is a standards-based process assessment method by which organisations can self-assess their processes in a transparent and efficient manner using a decision support system (DSS).

The next section discusses current literature on ITSM process assessment and overviews the international standards, DSS and the Process Virtualisation Theory that are used in our study. This is followed by a description of the Design Science Research (DSR) methodology used. The subsequent sections present the artefact design and evaluation phases. We then discuss the role of international standards and DSS in the virtualisation of process assessment, and the contribution of our artefact towards CSI. Finally, we provide conclusions and implications for future research.

## 2  Literature Review

An IT service is typically delivered with a combination of people, processes and technology and it should be defined with agreed levels of services to customers (TSO 2011). The use of IT to support business processes is crucial in the differentiation of IT services from a conventional definition of service (Spath et al. 2011). Service improvement can be facilitated by the accumulation of individual process improvements in ITSM. In the evaluation of software quality, it is recognised that assessing and improving a process is a means to improve product quality, and evaluating and improving product quality is one means of improving the system quality (ISO/IEC 2011b). In the ITSM context, this can be recognised as assessment of a process is a means to improvement, and evaluation and improvement of ITSM processes is one means of improving IT service quality as a whole. The ITIL framework supports this notion by presenting a service lifecycle with a continual improvement approach (ISO 2012). We present an overview of existing ITSM process assessment methods next, followed by an overview of the international standards and the DSS technology used to build our research artefact. Finally we introduce Process Virtualisation Theory which is later revisited to discuss virtualisation of ITSM process assessments.

### 2.1 Existing ITSM Process Assessment Methods

The potential of the ISO/IEC 15504 standard beyond its original software engineering focus has been reported (Coletta 2007; Rout et al. 2007) with claims that the ISO/IEC 15504 standard can be the "silver bullet as a centre of several extensions, if the extending standards can be structured in processes" (Malzahn 2009). A standard approach provides the objectivity required to measure process improvements effectively (Hilbert and Renault 2007). In response to increasing interest in the application of the standard, Mesquida et al. (2012) executed a systematic literature review of ITSM process improvement based on ISO/IEC 15504 and found 28 relevant primary studies. One is linked to the ITSM international standard, ISO/IEC 20000 (Nehfort 2007), whereas ten studies relate to the use of ITIL and ISO/IEC 15504. Using ITIL processes and ISO/IEC 15504, Barafort et al. (2002) provided evidence of repeatable and objective improvement in IT service quality. Extensive work on the combination of ITIL and ISO/IEC 15504 led to the development of an ITSM process assessment method called Tudor IT Process Assessment, or TIPA for ITIL (Barafort et al. 2009). Besides academic research, TIPA is also promoted as a commercial framework for ITSM process assessment (Renault and Barafort 2014). TIPA has gained support for continually improving ITSM processes (Barafort et al. 2014; Cortina et al. 2013; St-Jean 2009) and an approach to evaluate TIPA benefits to reduce assessment costs has been presented (St-Jean and Mention 2009). Furthermore, TIPA has been extended to present a service innovation framework in ITSM (Barafort and Rousseau 2009).

ITSM process assessment methods are discussed as best practice guidelines in the IT industry. Many of the solutions offered for ITSM process assessment are commercially available and aimed at selling organisations either a self-assessment toolkit or providing consultancy services as part of improvement initiatives, for example, TIPA for ITIL (Barafort et al. 2009); SPICE 1-2-1 (Nehfort 2007); SCAMPI using CMMI-SVC (CMMI 2011) and IT service CMM (Clerc and Niessink 2004). Other approaches emerged from industry best practice, particularly from ITIL (AXELOS 2014; MacDonald 2010). The measurement frameworks of ITSM process assessment methods are based on one of two models: CMM/CMMI and ISO/IEC 15504. ITIL is the most commonly used process reference model for ITSM process assessment. Non-ITIL approaches such as CMMI for Services (CMMI 2010) or eSCM for service providers (Hyder et al. 2004) also provide transparent models for assessment.

### 2.2 International Standards

#### 2.2.1 ISO/IEC 20000

The International Organisation for Standardisation (ISO) has developed requirements and guidance for ITSM in the form of the ISO/IEC 20000 standard. The standard has undergone a number of updates and is currently synchronised with the latest ITIL 2011 edition (ISO/IEC 2011a). ISO/IEC 20000 specifies requirements for IT service providers to develop and improve a service management system (ISO/IEC 2012). A process reference model for the assessment of ITSM processes is Part 4 of the standard "that represents process elements in terms of purpose and outcomes" (ISO/IEC 2010). The reference model provides the key indicators to achieve the overall objectives of an ITSM process.

#### 2.2.2 ISO/IEC 15504

ISO/IEC 15504 is the international standard for process assessment. It defines six process capability levels (CL0 to CL5): CL0 – Incomplete process; CL1 – Performed process; CL2 – Managed process;

CL3 – Established process; CL4 – Predictable process; and CL5 – Optimising process. CL0 suggests a lack of effective performance of the process. At CL1, a single process attribute is defined. There are two specific process attributes defined for each of the other process capability levels. Therefore a total of nine process attributes (PA1.1 to PA5.2) exist in the measurement framework. At a more granular level, a number of explicit process indicators are defined for each process attribute. These process indicators provide criteria to assess process capability in finer detail (ISO/IEC 2004a). Process assessment is conducted in a standard manner when it is compliant with ISO/IEC 15504-2 requirements where the assessors collect objective evidence against process indicators to determine process capability (ISO/IEC 2004b).

Beyond the software engineering discipline, the ISO/IEC 15504 standard is established as a general process assessment standard and is being transformed into a new standard family of ISO/IEC 330xx series (Jung et al. 2014). The fundamental evolution of the ISO/IEC 15504 standard architecture has attracted the interest of other industry sectors (Cortina et al. 2014). Some of the widely recognised projects to extend the use of ISO/IEC 15504 include Automotive SPICE, SPICE for Space, Enterprise SPICE, Banking SPICE and MediSPICE (Cortina et al. 2014; Van Loon 2007).

## 2.3 Decision Support System

Although traditionally associated with strategic decision-making for managers (Alter 1980), DSS is a general term for any information system that supports decision-making activities of individuals and groups (Power et al. 2011). A DSS presents the opportunity to eliminate the need for subjective judgment to determine process capability levels and provide process improvement recommendations in the SMPA approach. A knowledge-driven DSS that suggests or recommends actions to managers is highly relevant to our research. Such DSS can use technological rules and knowledge bases in which "knowledge" is stored in the form of rules. Knowledge-driven DSS uses an inference engine to process rules or identify relationships in data. Moreover, DSS enables specialised problem-solving based on the knowledge about a particular domain (Power et al. 2011). The DSS in the SMPA approach stores knowledge items of process improvements based on the ITIL framework. The DSS facilitates understanding of problems since low process capability scores indicate process risks. The DSS helps process managers make decisions to mitigate process risks and commence process improvement initiatives.

Our review found only one approach (Nehfort 2007) that reported the use of a software tool to conduct ITSM process assessments while only a handful of other tools were discussed in the literature. The software tools were designed to be used by the assessor in rating process attributes. While a software tool could minimise paper handling and manual work, it did not significantly impact the entire method of ITSM process assessment. In other words, the existing assessment tools may qualify as communications-driven, data-driven or document-driven DSS; however they cannot be classified as knowledge-driven DSS due to the lack of technological rules and knowledge base to recommend actions to process managers.

## 2.4 Process Virtualisation Theory

The Process Virtualisation Theory (PVT) developed by Overby (2008) is designed to explain whether any process is suitable to be followed virtually or not, i.e. the virtualisability of a process. Process virtualisation is a recent IS trend as seen in virtualisation of friendship using social networking sites, virtualisation of shopping via e-commerce or virtualisation of education using online learning platforms (Bose and Luo 2011). According to PVT, there are four requirements that have a negative relationship with process virtualisability. The requirements are: (a) sensory requirements – process stakeholders enjoy sensory experience of the process; (b) relationship requirements – process stakeholders interact with each other; (c) synchronisation requirements – efficient operation of process activities; and (d) identification and control requirements – process activities require unique identification of process stakeholders and control of its actions (Overby 2008). The theory also posits three IT-enabled moderating factors: (a) representation; (b) reach; and (c) monitoring capabilities that enable virtual processes. We use the three factors in order to articulate the SMPA approach later in the *Discussion* section.

# 3   Methodology

We designed and evaluated an ITSM process assessment method to address the stated research problem. We used an iterative design process to develop the SMPA approach and interpretative case studies to evaluate the usability of the SMPA approach. We followed the Design Science Research (DSR) methodology (Gregor and Jones 2007; Hevner et al. 2004; Peffers et al. 2007) in our research. DSR in IS has been used most commonly for generating field-tested and theoretically-grounded knowledge (McLaren et al. 2011). Our research artefact is a method for ITSM process assessments based on the international standards and implemented using a DSS. The artefact design elements and evaluation activities that we undertook in this research are illustrated in Figure 1.

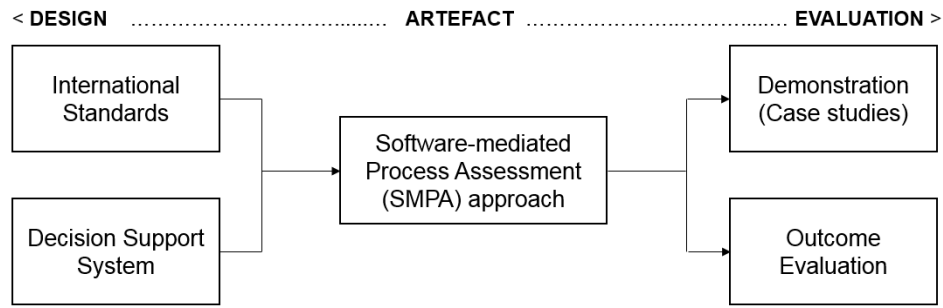

*Figure 1: Artefact Design Elements & Evaluation Activities*

Some of the most challenging problems of IS research are research relevance and practical utilisation (Carlsson 2007). The DSR methodology proposes that the output of research activities should provide practical design knowledge based on field-tested and grounded technological rules (Carlsson 2007). A "technological rule" is a prescription to follow if one wants to achieve a stipulated outcome in a standard setting (Bunge 1967). A "heuristic" form of technological rule can be designed in a typical qualitative format: "If you want to achieve Y in situation Z, then perform *something like* [emphasis added] action X" (Van Aken 2004). The SMPA approach is akin to a set of heuristic technological rules to develop a novel and practical method for ITSM process assessments. The design of the SMPA approach is discussed in detail next. The design process that was followed to develop the SMPA approach has been previously reported (Shrestha et al. 2014).

## 4 Artefact Design

Our research artefact, the SMPA approach, being software-mediated, uses a DSS to automate and virtualise the ITSM process assessment activities. In this section, we describe the phases of the SMPA approach, including the theoretical justification of the activities in each phase. Table 1 lists the four phases of the SMPA approach.

| Phase | DSS Functionality | Description |
| --- | --- | --- |
| Phase 1 Preparation | Process selection method | Define assessment goals, context and scope |
| Phase 2 Survey | Online survey | Collect responses to explicit assessment questions directly from participants |
| Phase 3 Measurement | Process capability rating | Analyse responses transparently to measure process capability |
| Phase 4 Improvement | Knowledge base | Use assessment results to guide process improvement |

*Table 1. Phases in the Software-mediated Process Assessment (SMPA) approach*

The first phase is preparation. In this phase, information about organisation profile, processes to assess and assessment participants along with their process roles are captured using the DSS. Each participant belongs to one of the three roles for any process: process manager, process performer or external process stakeholder. The second and third phases survey the process stakeholders and then measure process capability based on the survey responses according to the ISO/IEC 15504 standard. The final phase generates an assessment report that recommends process improvements. With the application of the SMPA approach, organisations can focus on the process improvement efforts rather than being concerned about the method and cost of repeated process assessments. A detailed architecture of the SMPA approach is illustrated in *Figure 2*. The four phases are discussed in detail next.

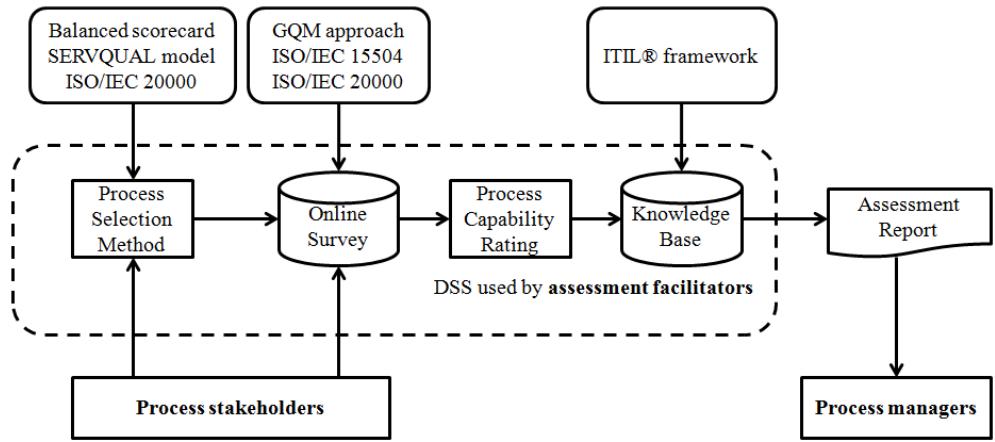

*Figure 2: The SMPA Architecture*

## 4.1 Process Selection Method

The process selection method was guided by the principles of the Balanced Scorecard (Kaplan and Norton 1992) and the SERVQUAL model (Parasuraman et al. 1985). With the input from the process stakeholders, the DSS assists in the selection of critical processes based on business drivers and stakeholders' service gap perceptions. Comprehensive details of the process selection method have previously been reported (Shrestha et al. 2015b). Four ITSM processes: Service Level Management; Change Management; Problem Management; and Configuration Management were selected as candidate processes for assessment.

## 4.2 Online Survey

While existing ITSM process assessments rely on process-specific indicators that demonstrate objective evidence of process capability, the SMPA approach facilitates a top-down approach where each ITSM process is defined with a goal and then assessment is guided by explicit questions and metrics that are set to goal attainment. The structure of the survey questionnaire is guided by the Goal-Question-Metric (GQM) approach (Basili et al. 2002). Following the GQM approach, assessment questions for the survey were generated by analysing all standard indicators to construct singular, fine-grained and close-ended assessment questions. A total of 46 questions specific to the four ITSM processes at capability level 1 (PA1.1) and 127 general questions for all processes at capability levels greater than 1 (PA2.1 to PA5.2) were generated.

The DSS is designed to collect quality data for measurement. Using the DSS, the responsibility to provide information about process capability is transferred to the process stakeholders. This shift from the current practice where assessors are responsible to collect assessment data means that with the SMPA approach, the assessors do not need to conduct interviews and make subjective judgments on process capability. For example, an assessor's open-ended question for the problem management process based on the standard practice "RES.3.1 Identify problems" could be "Can you tell me how you record the problems?" By comparison, the assessment question in the survey is formed as "Do you know if identified problems are recorded?" in a close-ended format, so that the assessment facilitator can analyse survey responses objectively based on a concrete set of answer options.

## 4.3 Measurement

The assessment questions were grouped to determine process capability levels 1-5 and every question was designed to have consistent answer options using the rating scale: Not (N), Partially (P), Largely (L) and Fully (F) – also referred as the NPLF scale – as defined in the measurement framework of the ISO/IEC 15504 standard. This rating is a knowledge metric to capture what ITSM process stakeholders know about the process. Rather than the assessment team making a subjective choice of the indicator rating, the SMPA approach uses this metric to collect and objectively measure feedback from the process stakeholders directly.

The ISO/IEC 15504-2 requirements are used for the calibration of process attribute ratings. According to the measurement framework in the standard, a particular capability level can be achieved if a process meets two conditions: (a) the target level is fully or largely achieved, i.e. the rating of "Fully" or "Largely" for the process attributes at that level; and (b) the lower levels are fully achieved, i.e. the rating of "Fully" for all lower level process attributes. For example, a process can only achieve CL3 if it obtains a "Fully" or "Largely" score in PA3.1 and PA3.2 and all process attributes below CL3 (i.e. PA1.1, PA2.1 and PA2.2) must be "Fully" achieved. All responses contribute equal weight to each assessment question. However, responses are implicitly weighted according to how the process roles are allocated to the assessment questions as the number of questions differs according to the process roles, and this will subsequently affect the process capability score. The final score of each process attribute is determined by calculating the arithmetic mean value of all the responses using the scale percentage based on the ISO/IEC 15504 standard measurement framework. The DSS in the SMPA approach also computes the coefficient of variation to determine reliability in terms of the spread of responses.

## 4.4 Improvement

After each process questionnaire was formulated, knowledge items were generated for all questions based on the best practice guidelines of the ITIL framework. A knowledge item for each question is extracted from the knowledge base and compiled in the assessment report when the normalised mean of all responses to the question – referred to as the knowledge score for the question – demonstrates risks (i.e. a knowledge score of Not or Partially).

For every assessment question, two components – observation and recommendation – are combined to generate a process improvement knowledge item. The observation component of a knowledge item lists the current state of the process capability. Likewise the recommendation component of a knowledge item is based on the best practice guidelines from the ITIL framework to achieve higher capability levels. For instance, if a question asked "Do you know if X is performed?" the associated knowledge item may consist of two components: (a) Observation: "X is not performed well"; and (b) Recommendation: "According to ITIL, Y can be considered to perform X well". For all 173 assessment questions developed in the SMPA approach, 151 corresponding knowledge items were developed to address risks associated with the process in question.

In the SMPA approach, the use of a DSS can automate (a) assessment data collection using online surveys, (b) data analysis to calculate process capability scores, and (c) reporting from a context-based knowledge base of process improvement recommendation items. These opportunities translate to significant cost savings through avoidance of the use of costly assessors and consultants while enabling self-assessments for IT organisations with fast turnaround time. The SMPA approach was evaluated at two IT service providers to determine its usability to process managers for decision making. Evaluation results are discussed next.

# 5 Artefact Evaluation

Because evaluation based on the actual decision quality is time consuming and difficult to measure, soft measures such as perceived decision quality factors have been used in DSS research (Jarupathirun and Zahedi 2007). Perceived decision quality and efficiency measure perception after the decision has been made whereas expected decision quality and efficiency can be evaluated prior to making decisions (Parikh et al. 2001). Perceived decision quality and efficiency have been used to explore successful use of a web-based spatial DSS (Jarupathirun and Zahedi 2007) and other web-based DSS (e.g. Gu and Wang 2009). Due to temporal constraints, expected decision quality and expected decision efficiency were used for evaluation of the SMPA report.

The SMPA approach was implemented at the two IT service providers in October 2013. Assessment data collection was completed by early November 2013. The SMPA reports were emailed to the assessment facilitators in early December 2013. After receiving confirmation from the assessment facilitators that the SMPA report has been reviewed by the relevant process managers at both organisations, in-depth interviews were conducted with relevant process managers in February 2014 to evaluate their expectations on the usability of the SMPA report. We used four usability characteristics derived from the international standard for software quality evaluation (ISO/IEC 2011b) to evaluate the SMPA report. The four characteristics were effectiveness, efficiency, usefulness and trust in relation to the expected usability of the SMPA report.

We asked five ITSM process managers (coded: "MgrA" & "MgrB" at Organisation 1; and "MgrC", "MgrD" & "MgrE" at Organisation 2) if the SMPA report is useful to make process improvement decisions. The interviews were recorded and transcribed. We analysed and coded qualitative data sourced from the interview transcripts. We marked ☑ where we found that a particular usability characteristic was strongly supported by the process manager. Likewise ☒ indicated that the usability characteristic was not strongly supported. Evaluation results are summarised in *Table 2*. Full evaluation work along with details of in-depth interviews have been reported in the past (Shrestha et al. 2015a).

| Usability characteristic | Case evidence | Selected key comments |
|---|---|---|
| Effectiveness | ☒ ☑ ☑ ☑ ☑ | ☒ MgrA: I've misunderstood the report … the report wasn't clear … I am trying to learn to read the report. |
| | | ☑ MgrC: … my decision is based on accurate information and hence will be a correct decision with this [SMPA] report… |
| | | ☑ MgrD: the answers that have come out [of the SMPA report] seem to be a far more accurate assessment of our environment. |
| Efficiency | ☒ ☒ ☒ ☒ ☒ | ☒ MgrA: when I went through it [SMPA report], it seemed to overcomplicate Problem Management [process]… It is really hard and time consuming to read |
| | | ☒ MgrB: it [SMPA report] probably would take longer to read … they're too broad and there may be a lot of stuff to read through … |
| | | ☒ MgrC: …because I must admit, the first time I looked at it [SMPA report], I was overwhelmed. This is a lot of detail and its 35 pages long! How am I going to do this? |
| Usefulness | ☑ ☑ ☑ ☑ ☑ | ☑ MgrB: Yes… its useful … it has a market in terms of if someone wants to get an idea of improvement |
| | | ☑ MgrC: we've already gone through some areas of the report and looked at areas where we need to improve… |
| | | ☑ MgrE: It's useful for showing us the subject areas for where our next steps are … |
| Trust | ☑ ☑ ☑ ☑ ☑ | ☑ MgrB: the online one [SMPA report] is going to be more reliable because you've got a broader audience and the same assessment criteria and formula happening. |
| | | ☑ MgrC: [SMPA report] is a truer representation of where the organisation is at, with respect to its process maturity. |
| | | ☑ MgrE: Between me and the two other people I spoke to, I think we did pretty much come to a consensus trusting the results we got from this [SMPA report]. |

*Table 2. Summary of Evaluation Results of SMPA Report at two IT service providers*

One of the most significant evaluation findings is that most process managers expected that better quality decisions could be made on process improvements. It was also found that the process managers considered the expected utility and trust of the SMPA report to be highly positive. Process managers thought the SMPA report is time consuming to read and implement.

In response to negative expected decision efficiency for the SMPA report, the structure and content of the SMPA report can be modified for clarity. Changes in the report template, presentation of assessment results and listing of process improvement recommendations have been suggested to address the shortcomings of the SMPA report. Hence, further work is planned to make the SMPA report succinct and targeted to the main audience of the report – the process managers. The report must provide clear rationale and directions to the process managers to implement process improvements. We discuss the design knowledge that emerged from our research next.

## 6 Discussion

The design knowledge for the SMPA approach satisfies many of the criteria for partial, nascent theory (Gregor and Hevner 2013). We used the DSR knowledge contribution framework (Gregor and Hevner 2013) to discuss our research knowledge contributions. The DSR knowledge contribution framework presents two dimensions based on the existing state of knowledge in both the problem and solution domains. The problem domain is represented by the challenges of ITSM process assessment. The solution domain is represented by the international standards and DSS capabilities. Using the DSR contribution types presented by Gregor and Hevner (2013), Level 1 and Level 2 contributions are evident in this research. At level 1, situated implementation was constructed as a DSS for the SMPA approach. Likewise a more general artefact in the form of a method (SMPA approach) is proposed as the level 2 contribution. The design knowledge in this research, however, has not yet evolved to the stage where it could be termed "design theory", i.e. Level 3 contribution (Gregor and Hevner 2013). We attributed the successful evaluation results of the SMPA approach to two key design principles. The two design principles are discussed next, followed by the role of the SMPA approach in the virtualisation of ITSM process assessment and for CSI.

### 6.1 Role of the International Standards

International standards harmonise technical specifications of products and services by offering transparent benchmarks (Marquardt and Juran 1999). Even though standards provide authoritative statements of good professional practice, such statements are general principles rather than specific activities (Bevan 2001). Due to this role of the international standards, they promote transparency in the way activities are undertaken. The SMPA approach provides prescriptive details of activities to be undertaken for ITSM process assessment. However the artefact is scaffolded by the principles of international standards in order to support and validate the prescribed activities. In this light, the SMPA approach follows the international standards of ITSM and process assessment to transparently conduct ITSM process assessments. The use of the international standards in the design of the artefact promotes quality improvement, cost savings and increases in productivity and competitive advantage (ISO 2015).

Standards have been credited with facilitating communication in IS and making the discipline more consistent (Getronics 2006). The true value of a standard evolves by facilitation of data exchange and consequently reduction in the cost of information. Quality and cost efficiency are two major objectives in almost all best practice standards (ISO 2001). Therefore standards should belong to the public domain and be universally applicable in order to be used in a transparent manner (Kumbakara 2008). The ISO/IEC 15504 standard (ISO/IEC 2004a) mandates the requirement of a documented assessment process that helps to determine the workflow for ITSM process assessments. Following this standard, the SMPA approach provides a transparent method to conduct assessments.

### 6.2 Utility of the DSS

The assessment data collection and validation, rating of the process capability and reporting of the assessment results require ITSM information to be gathered, aggregated, evaluated and presented. Therefore, having a sound information processing capability is an important requirement for the SMPA approach. In this scenario, the DSS for the SMPA approach can be a cost effective solution. The DSS can store and analyse data sets from several iterations of targeted stakeholder responses of assessment questions. In this way data analysis can be low cost and happen in real time for each assessment. Moreover, DSS can extend the bounds of rationality for decision makers through their capabilities (Todd and Benbasat 1999).

The automatic storage of collected information provides an opportunity for validated data to be used to compare process assessment results for benchmarking and demonstration of process improvement. This is important as currently no aggregated analysis could be carried out with the existing manual process assessment methods. While there are software tools available for assessors to input assessment data, no software tools have been reported that can capture information directly from the process stakeholders and analyse the collected assessment responses using the international standard for process assessment. This feature is implemented in the DSS employed by the SMPA approach.

### 6.3 Virtualisation of ITSM Process Assessment

The most prominent themes emerging from our research are the role of international standards and the utility of DSS technology to automate ITSM process assessments. With these two design elements, ITSM process assessments can be "virtualised", i.e. absence of physical interaction between people, for instance in the context of virtual teams (Fiol and O'Connor 2005). The impact of the SMPA approach in ITSM process assessments can be observed from the lens of PVT. The SMPA approach is supported by the three features promoted by PVT to enable virtualisation of ITSM process assessment, viz. representation; reach; and monitoring capabilities. How these three features are used as pre-design criteria to develop and justify the SMPA approach is discussed next.

#### 6.3.1 Representation in the SMPA Approach

In terms of "representation", the SMPA approach represents standard process information for assessment. The ITIL best practice framework and the international standards for ITSM and Process

Assessment are represented in the SMPA approach to facilitate transparency in the way process assessments are conducted virtually. Without a DSS, compilation of an assessment report with process improvement recommendations would require an assessment team with multi-disciplinary skills and expertise in process assessment and ITSM, working for a considerable period of time to compile relevant recommendations. The DSS can efficiently draw upon expert knowledge of process improvements from its knowledge base, thus virtualise expert representations of ITSM best practices. With the use of the online survey for assessment data collection and a knowledge base to compile the process improvement report, the SMPA approach allows the entire ITSM process assessment workflow to be executed electronically. Therefore, virtualised ITSM process assessment enabled by the SMPA approach represents the entire assessment experience with consistent and transparent activities throughout the process improvement journey.

### 6.3.2 Reach in the SMPA Approach

According to PVT, IT can increase "reach" to engage more process stakeholders in less time and effort (Overby 2008). The SMPA approach can represent the assessment results from the entire population of process stakeholders. With an online survey interface, the SMPA approach can query and capture responses from process participants regardless of geography, thereby offering a wider "reach". Use of online surveys in psychological studies has been linked with efficiency due to automation that also enables expansion of the scale and scope of such studies (Kraut et al. 2004). Moreover, online surveys can gather credible data input even from the introverts in an organisation who respond best in quiet environments as discussed by Cain (2013). Online surveys are also ideally suited for remote data collection from a global IT workforce as compared to document reviews or interviews. The prevalent growth of outsourcing of IT service functions and the use of virtual IT teams across the globe means that online surveys can be a suitable assessment data collection tool to perform ITSM process assessments, allowing synchronous participations from different locations. Broader participation yields a comprehensive coverage of assessment feedback that is not feasible in manual assessments.

Besides reaching the wide cross-section of process stakeholders, the SMPA approach can also capture the depth of responses since online surveys help process stakeholders provide granular and detailed feedback. Using the online survey, the responses from the process assessment exercise can be grouped in different process roles, thereby making it possible to analyse scenarios such as when process managers provide a skewed opinion of the process being performed in contrast with the process performers. Such readings can help IT service managers to perform gap analysis and understand deficiencies in the process activities. These types of analysis are feasible to solicit from online surveys but would not be easy to realise from assessment interviews.

### 6.3.3 Monitoring Capability in the SMPA Approach

Based on "monitoring capability", the DSS in the SMPA approach can solicit responses from the process stakeholders and track their assessment progress. This is perhaps the most significant value of the SMPA approach in terms of virtualisability of ITSM process assessments. Using the monitoring capability of the DSS, assessment responses can be verified and analysed. The SMPA approach supports enhanced ability to track assessment participation and access granular process improvement recommendations. Likewise, the ability to store historical data on process performance means that the virtualised SMPA approach is ideal for repetitive and formative self-assessments.

The logic of process capability determination and calculation of the reliability score of the survey responses is a feature of the SMPA approach that is not explicitly stated in the ISO/IEC 15504 standard. This is an example of how the functionality of the SMPA approach could be expanded and use several data analysis techniques to develop an objective measure of process capability without the need of discussion among the assessment team members. The SMPA approach can leverage its monitoring capability to process these calculations in a more consistent manner than humans, thereby supporting virtualisation.

Anecdotal evidence suggests that manually entering data and subjective judgment based on interviews and document reviews can be error-prone and requires a longer time commitment from the assessment team. Consequently the entire process assessment method becomes subjective and costly. This means that repeated process assessments to build a repository of process improvement recommendations are unlikely to be given a priority due to the significant workload involved in the process assessment effort itself. The SMPA approach can monitor the entire assessment cycle in a virtual setting that can eliminate latency for process improvement efforts.

## 6.4 SMPA Approach for Continual Service Improvement

The Theory of Constraints (Goldratt and Cox 1992) suggests that the continuous improvement principle cannot be solely determined by improving processes without understanding the interactions of the processes as a system (Dettmer 1997). However, the measurement of processes for improvement is a requirement to facilitate service improvements (Cannon 2011). If the process assessment activities are not supported by a commitment to improve processes, then the IT service organisations will only have a system to identify the problems but they will not have any support for service improvement (Malzahn 2009). Therefore, an ideal application of the SMPA approach is within an environment that provides initial assessment before continuous improvement opportunities with checkpoint assessments for review. This principle has been prominently discussed not only within the ITSM discipline but also in other quality disciplines such as continuous improvement methods in Total Quality Management (TQM) (Powell 1995) and continuous improvement in the ISO 9000 standard (Marquardt and Juran 1999).

One of the key principles of TQM suggests that process deficiencies are the root cause of most of the mistakes made by individuals in organisations. By improving the processes, repetition of such mistakes can be prevented (Gilbert 1992). In order to improve processes, ongoing assessments are a requirement for CSI in the ITSM discipline (Lloyd 2011). According to the continuous improvement literature, organisations can only advance to a new level after an earlier status has been achieved (Bessant and Caffyn 1997). Such an incremental, step-by-step improvement approach is consistent with the views of CSI where ITSM organisations review their past decisions and make better decisions through gradual process improvements.

Process improvement activities require periodic process assessments (Malzahn 2009). The approach of conducting periodic assessment for process improvement has been reported in the field of software process improvement for small firms (Cater-Steel et al. 2005) and project management (Malzahn 2009). Likewise, the SMPA approach is focused on process assessment; however it is important to understand the significance of repeated ITSM process assessments for CSI. Since process improvement can be measured through repeated assessments, self-assessment of ITSM processes in a virtual setting presents is an opportunity for IT service providers to propel CSI.

## 7 Conclusion

The SMPA approach demonstrated the application of software mediation to bring transparency and efficiency to the way process assessments are conducted. Transparency issues in ITSM process assessment were addressed by following a goal-oriented measurement of ITSM processes using an international standard. Besides the use of the international standard for process assessment, the virtualisation of the ITSM process assessment is supported by two features: (a) online surveys to allow faster and consistent assessment data collection and analysis; and (b) knowledge base for process improvement recommendations from the ITIL library. The virtualised SMPA approach enables IT service organisations to self-assess the capability of their ITSM processes.

The case study in this research included certain limitations. First, regarding internal validity, evaluation data were collected using qualitative research methods in two case study organisations. A recognised limitation of the qualitative case study approach is the lack of ability to generalise the findings. Although the artefact can provide an objective assessment, the assessment results are still based on the responses of the process stakeholders. Despite the innovative prospects of our research, it is necessary to conduct comprehensive evaluations of ITSM process assessments for further improvement of the artefact. In order to obtain a richer view of integration of the SMPA approach, we intend to apply the artefact in other organisations and with more processes in order to confirm and generalise the applicability and effectiveness of the SMPA approach. Future research could explore feedback cycles from several design-evaluation iterations. This should lead to a robust method defined as a design theory (Gregor and Jones 2007) or a process theory (Markus and Robey 1988) capable of virtualising process assessments in ITSM.

In summary, the SMPA approach provides a new opportunity for virtualisation in the way process assessments are conducted in IT organisations. Beyond the discipline of ITSM, the SMPA approach can potentially be applied to other domains. For example, COBIT has released an ISO/IEC 15504 compliant assessment model for its IT governance processes (ISACA 2015). With the expanding significance and reach of the ISO/IEC 15504 standard and the soon-to-be-published ISO/IEC 330xx series, the SMPA approach is expected to be a useful virtual method for process assessments in other disciplines beyond ITSM.

## 8 References

Alter, S.L. 1980. *Decision Support Systems: Current Practice and Continuing Challenge*. Reading, MA, USA: Addison-Wesley.

AXELOS. 2014. "ITIL® Maturity Model."   Retrieved 30 Jun, 2015, from http://www.axelos.com/itil-maturity-model

Barafort, B., Betry, V., Cortina, S., Picard, M., St-Jean, M., Renault, A., and Valdès, O. 2009. *ITSM Process Assessment Supporting ITIL*. Zaltbommel, Netherlands: Van Haren Publishing.

Barafort, B., Di Renzo, B., and Merlan, O. 2002. "Benefits Resulting from the Combined Use of ISO/IEC 15504 with the Information Technology Infrastructure Library (ITIL)," *4th International Conference on Product Focused Software Process Improvement*, London, UK: Springer-Verlag, pp. 314-325.

Barafort, B., and Rousseau, A. 2009. "Sustainable Service Innovation Model: A Standardized IT Service Management Process Assessment Framework," in: *Software Process Improvement*. Germany: Springer Berlin Heidelberg, pp. 69-80.

Barafort, B., Rousseau, A., and Dubois, E. 2014. "How to Design an Innovative Framework for Process Improvement? The TIPA for ITIL Case," in: *Systems, Software and Services Process Improvement*. Germany: Springer Berlin Heidelberg, pp. 48-59.

Basili, V.R., Caldiera, G., Rombach, H.D., and van Solingen, R. 2002. "Goal Question Metric (GQM) Approach," *J. Marciniak: Encyclopedia of Software Engineering* (1), pp 578-583.

Bessant, J., and Caffyn, S. 1997. "High-involvement Innovation Through Continuous Improvement," *International Journal of Technology Management* (14:1), pp 7-28.

Bevan, N. 2001. "International Standards for HCI and Usability," *International Journal of Human-Computer Studies* (55:4), pp 533-552.

Bose, R., and Luo, X. 2011. "Integrative Framework for Assessing Firms' Potential to Undertake Green IT Initiatives via Virtualization–A Theoretical Perspective," *Journal of Strategic Information Systems* (20:1), pp 38-54.

Bunge, M. 1967. *Scientific Research 2: The Search for Truth*. Berlin, Germany: Springer.

Cain, S. 2013. *Quiet: The Power of Introverts in a World That Can't Stop Talking*. NY, USA: Broadway Paperbacks.

Cannon, D. 2011. *ITIL Service Strategy*. London, UK: The Stationery Office.

Carlsson, S.A. 2007. "Developing Knowledge Through IS Design Science Research," *Scandinavian Journal of Information Systems* (19:2), pp 75-86.

Cater-Steel, A., Toleman, M., and Rout, T. 2005. "An Evaluation of the RAPID Assessment-based Process Improvement Method for Small Firms," *9th International Conference on Evaluation and Assessment in Software Engineering*, Keele, UK.

Clerc, V., and Niessink, F. 2004. *IT Service CMM: A Pocket Guide*. Zaltbommel, Netherlands: Van Haren Publishing.

CMMI. 2010. "CMMI® for Services, Version 1.3," Software Engineering Institute, Carnegie Mellon University, MA, USA.

CMMI. 2011. "Standard CMMI® Appraisal Method for Process Improvement (SCAMPI) A, Version 1.3: Method Definition Document," Software Engineering Institute, Carnegie Mellon University, MA, USA.

Coletta, A. 2007. "An Industrial Experience in Assessing the Capability of Non-software Processes Using ISO/IEC 15504," *Software Process: Improvement and Practice* (12:4), pp 315-319.

Cortina, S. 2010. "Why Perform Process Assessments?"   Retrieved 31 Jul, 2015, from http://www.itsmportal.com/columns/why-perform-process-assessments

Cortina, S., Mayer, N., Renault, A., and Barafort, B. 2014. "Towards a Process Assessment Model for Management System Standards," in: *Software Process Improvement and Capability Determination*. Switzerland: Springer International Publishing, pp. 36-47.

Cortina, S., Renault, A., and Picard, M. 2013. "TIPA Process Assessments: A Means to Improve Business Value of IT Services," *International Journal of Strategic Information Technology and Applications* (4:4), pp 1-18.

Dettmer, H.W. 1997. *Goldratt's Theory of Constraints: A Systems Approach to Continuous Improvement*. Milwaukee, USA: ASQ Quality Press.

Dyke, T.P., Prybutok, V.R., and Kappelman, L.A. 1999. "Cautions on the Use of the SERVQUAL Measure to Assess the Quality of Information Systems Services," *Decision Sciences* (30:3), pp 877-891.

England, R. 2012. "Why Process Maturity is a Useless Metric for ITSM Improvement."   Retrieved 29 Jul, 2015, from http://www.itskeptic.org/content/why-process-maturity-useless-metric-planning-improvement

Fayad, M.E., and Laitinen, M. 1997. "Process Assessment Considered Wasteful," *Communications of the ACM* (40:11), pp 125-128.

Fiol, C.M., and O'Connor, E.J. 2005. "Identification in Face-to-face, Hybrid, and Pure Virtual Teams: Untangling the Contradictions," *Organization Science* (16:1), pp 19-32.

Getronics. 2006. *Implementing Leading Standards for IT Management*, (1st ed.). Zaltmobbel, Netherlands: Van Haren Publishing.

Gilbert, G.R. 1992. "Quality Improvement in a Federal Defense Organization," *Public Productivity & Management Review* (16:1), pp 65-75.

Goldratt, E.M., and Cox, J. 1992. *The Goal: A Process of Ongoing Improvement*. MA, USA: North River Press.

Gregor, S., and Hevner, A.R. 2013. "Positioning and Presenting Design Science Research for Maximum Impact," *MIS Quarterly* (37:2), pp 337-355.

Gregor, S., and Jones, D. 2007. "The Anatomy of a Design Theory," *Journal of the Association for Information Systems* (8:5), pp 312-335.

Gu, L., and Wang, J. 2009. "A Study of Exploring the "Big Five" and Task Technology Fit in Web-Based Decision Support Systems," *Issues in Information Systems* (10:2), pp 210-217.

Hevner, A.R., March, S.T., Park, J., and Ram, S. 2004. "Design Science in Information Systems Research," *MIS Quarterly* (28:1), pp 75-105.

Hilbert, R., and Renault, A. 2007. "Assessing IT Service Management Processes with AIDA– Experience Feedback," *14th European Conference for Software Process Improvement*, Potsdam, Germany.

Hubbard, D.W. 2010. *How to Measure Anything: Finding the Value of "Intangibles" in Business*, (2nd ed.). New Jersey, USA: John Wiley & Sons.

Hyder, E.B., Heston, K.M., and Paulk, M.C. 2004. "The eSourcing Capability Model for Service Providers (eSCM-SP) v2, Part 1: Model Overview," Carnegie Mellon University, Pittsburgh, USA.

ISACA. 2015. "COBIT 5 Assessment Programme " Retrieved 28 Jul, 2015, from http://www.isaca.org/Knowledge-Center/cobit/Pages/COBIT-Assessment-Programme.aspx

ISO. 2001. "ISO Guide 72:2001 - Guidelines for the Justification and Development of Management System Standards." Geneva, Switzerland: International Organisation for Standardisation.

ISO. 2012. "Quality Management Principles." Geneva, Switzerland: ISO Central Secretariat.

ISO. 2015. "Benefits of International Standards." Retrieved 18 Jul, 2015, from http://www.iso.org/iso/home/standards/benefitsofstandards.htm

ISO/IEC. 2004a. "ISO/IEC 15504-2:2004 – Information Technology – Process Assessment – Part 2: Performing an Assessment." Geneva, Switzerland: International Organisation for Standardisation.

ISO/IEC. 2004b. "ISO/IEC 15504-3:2004 – Information Technology – Process Assessment – Part 3: Guidance on Performing an Assessment." Geneva, Switzerland: International Organisation for Standardisation.

ISO/IEC. 2005. "ISO/IEC 15504-4:2005 - Information Technology - Process Assessment - Part 4: Guidance on Use for Process Improvement and Process Capability Determination." Geneva, Switzerland: International Organisation for Standardisation.

ISO/IEC. 2010. "ISO/IEC TR 20000-4:2010 – Information Technology – Service Management – Part 4: Process Reference Model." Geneva, Switzerland: International Organisation for Standardisation.

ISO/IEC. 2011a. "ISO/IEC 20000-1:2011 – Information Technology – Service Management – Part 1: Service Management System Requirements." Geneva, Switzerland: International Organisation for Standardisation.

ISO/IEC. 2011b. "ISO/IEC 25010:2011 – Systems and Software Engineering – Systems and Software Quality Requirements and Evaluation (SQuaRE) - System and Software Quality Models." Geneva, Switzerland: International Organisation for Standardisation.

ISO/IEC. 2012. "ISO/IEC 20000-2:2012 – Information Technology – Service Management – Part 2: Guidance on the Application of Service Management Systems." Geneva, Switzerland: International Organisation for Standardisation.

Jarupathirun, S., and Zahedi, F. 2007. "Exploring the Influence of Perceptual Factors in the Success of Web-based Spatial DSS," *Decision Support Systems* (43:3), pp 933-951.

Jiang, J.J., Klein, G., and Carr, C.L. 2002. "Measuring Information System Service Quality: SERVQUAL from the Other Side," *MIS Quarterly* (26:2), pp 145-166.

Jung, H.-W., Varkoi, T., and McBride, T. 2014. "Constructing Process Measurement Scales Using the ISO/IEC 330xx Family of Standards," in: *Software Process Improvement and Capability*

*Determination,* A. Mitasiunas, T. Rout, R. O'Connor and A. Dorling (eds.). Springer International Publishing, pp. 1-11.

Juran, J.M., and Godfrey, A.B. 1999. *Juran's Quality Handbook*, (5th ed.). USA: McGraw-Hill.

Kane, D. 2012. "ITSM Maturity Assessments: A Value-based Approach."  Retrieved 15 Jul, 2015, from http://www.hazyitsm.com/2012/05/itsm-maturity-assessments-value-based.html

Kaplan, R.S., and Norton, D.P. 1992. "The Balanced Scorecard–Measures that Drive Performance," *Harvard Business Review* (70:1), pp 71-79.

Kettinger, W.J., and Lee, C.C. 1994. "Perceived Service Quality and User Satisfaction with the Information Services Function," *Decision Sciences* (25:5-6), pp 737-766.

Kettinger, W.J., and Lee, C.C. 2005. "Zones of Tolerance: Alternative Scales for Measuring Information Systems Service Quality," *MIS Quarterly* (29:4), pp 607-623.

Kraut, R., Olson, J., Banaji, M., Bruckman, A., Cohen, J., and Couper, M. 2004. "Psychological Research Online: Report of Board of Scientific Affairs' Advisory Group on the Conduct of Research on the Internet," *American Psychologist* (59:2), p 105.

Kumbakara, N. 2008. "Managed IT Services: the Role of IT Standards," *Information Management & Computer Security* (16:4), pp 336-359.

Lepmets, M., Mesquida, A.L., Cater-Steel, A., Mas, A., and Ras, E. 2014. "The Evaluation of the IT Service Quality Measurement Framework in Industry," *Global Journal of Flexible Systems Management* (15:1), pp 39-57.

Lloyd, V. 2011. *ITIL Continual Service Improvement*. London, UK: The Stationery Office.

MacDonald, I. 2010. "ITIL Process Assessment Framework." Manchester, UK: The Co-operative Financial Services.

Malzahn, D. 2009. "Assessing - Learning - Improving, an Integrated Approach for Self Assessment and Process Improvement Systems," *Fourth International Conference on Systems*, Cancun, Mexico, pp. 126-130.

Markus, M.L., and Robey, D. 1988. "Information Technology and Organizational Change: Causal Structure in Theory and Research," *Management Science* (34:5), pp 583-598.

Marquardt, D.W., and Juran, J.M. 1999. *The ISO 9000 Family of International Standards*. USA: McGraw-Hill.

McLaren, T.S., Head, M.M., Yuan, Y., and Chan, Y.E. 2011. "A Multilevel Model for Measuring Fit Between a Firm's Competitive Strategies and Information Systems Capabilities," *MIS Quarterly* (35:4), pp 909-929.

Mesquida, A.L., Mas, A., Amengual, E., and Calvo-Manzano, J.A. 2012. "IT Service Management Process Improvement based on ISO/IEC 15504: A Systematic Review," *Information and Software Technology* (54:3), pp 239-247.

Nehfort, A. 2007. "SPICE Assessments for IT Service Management according to ISO/IEC 20000-1," *The International SPICE 2007 Conference*, Frankfurt, Germany.

Overby, E. 2008. "Process Virtualization Theory and the Impact of Information Technology," *Organization Science* (19:2), pp 277-291.

Parasuraman, A., Zeithaml, V.A., and Berry, L.L. 1985. "A Conceptual Model of Service Quality and its Implications for Future Research," *Journal of Marketing* (49:4), pp 41-50.

Parikh, M., Fazlollahi, B., and Verma, S. 2001. "The Effectiveness of Decisional Guidance: An Empirical Evaluation," *Decision Sciences* (32:2), pp 303-332.

Peffers, K., Tuunanen, T., Rothenberger, M.A., and Chatterjee, S. 2007. "A Design Science Research Methodology for Information Systems Research," *Journal of Management Information Systems* (24:3), pp 45-77.

Peldzius, S., and Ragaisis, S. 2013. "Usage of Multiple Process Assessment Models," in: *Software Process Improvement and Capability Determination,* T. Woronowicz, T. Rout, R. O'Connor and A. Dorling (eds.). Germany: Springer Berlin Heidelberg, pp. 223-234.

PinkElephant. 2012. "PinkSCAN™ - Online Process Maturity Assessment."  Retrieved 12 May, 2012, from http://www.pinkelephant.com/Products/PinkONLINE/PinkScan/

Pitt, L.F., Watson, R.T., and Kavan, C.B. 1995. "Service Quality: a Measure of Information Systems Effectiveness," *MIS Quarterly* (19:2), pp 173-187.

Powell, T.C. 1995. "Total Quality Management as Competitive Advantage: A Review and Empirical Study," *Strategic Management Journal* (16:1), pp 15-37.

Power, D., Burstein, F., and Sharda, R. 2011. "Reflections on the Past and Future of Decision Support Systems: Perspective of Eleven Pioneers," in: *Decision Support,* D. Schuff, D. Paradice, F. Burstein, D.J. Power and R. Sharda (eds.). New York: Springer, pp. 25-48.

Praeg, C.-P., and Schnabel, U. 2006. "IT-Service Cachet - Managing IT-Service Performance and IT-Service Quality," *39th Annual Hawaii International Conference on System Sciences*, HI, USA: IEEE.

Renault, A., and Barafort, B. 2014. "TIPA for ITIL – From Genesis to Maturity of SPICE Applied to ITIL 2011," *European System & Software Process Improvement and Innovation*, Henri Tudor Institute, Luxembourg.

Rout, T.P., El Emam, K., Fusani, M., Goldenson, D., and Jung, H.W. 2007. "SPICE in retrospect: Developing a standard for process assessment," *Journal of Systems and Software* (80:9), pp 1483-1493.

Rudd, C., and Sansbury, J. 2013. "ITIL® Maturity Model and Self-assessment Service: User Guide," AXELOS Limited, Norwich, UK.

Shrestha, A., Cater-Steel, A., Tan, W.-G., and Toleman, M. 2014. "Software-mediated Process Assessment for IT Service Capability Management," *Twenty Second European Conference on Information Systems (ECIS 2014)*, Tel Aviv, Israel.

Shrestha, A., Cater-Steel, A., Toleman, M., and Rout, T. 2015a. "Evaluation of Software Mediated Process Assessments for IT Service Management," in: *Software Process Improvement and Capability Determination*. Springer International Publishing, pp. 72-84.

Shrestha, A., Cater-Steel, A., Toleman, M., and Tan, W.-G. 2015b. "A Method to Select IT Service Management Processes for Improvement," *Journal of Information Technology Theory and Application (JITTA)* (15:3), p 3.

Spath, D., Bauer, W., and Praeg, C.-P. 2011. "IT Service Quality Management: Assumptions, Frameworks and Effects on Business Performance," in: *Quality Management for IT Services- Perspectives on Business and Process Performance*. PA, USA: IGI Global, pp. 1-21.

St-Jean, M. 2009. "TIPA to keep ITIL going and going," *European System & Software Process Improvement and Innovation Conference*, Alcala de Henares, Spain.

St-Jean, M., and Mention, A.-L. 2009. "How to Evaluate Benefits of Tudor's ITSM Process Assessment?," *International SPICE Conference on Process Improvement and Capability dEtermination*, Turku, Finland.

Todd, P., and Benbasat, I. 1999. "Evaluating the Impact of DSS, Cognitive Effort, and Incentives on Strategy Selection," *Information Systems Research* (10:4), pp 356-374.

TSO. 2011. *The Official Introduction to the ITIL Service Lifecycle*. London, UK: The Stationery Office.

Van Aken, J.E. 2004. "Management Research Based on the Paradigm of the Design Sciences: The Quest for Field-tested and Grounded Technological Rules," *Journal of Management Studies* (41:2), pp 219-246.

Van Loon, H. 2007. *Process Assessment and ISO/IEC 15504: a reference book*, (2nd ed.). NY, USA: Springer-Verlag.

Walker, R.H., Johnson, L.W., and Leonard, S. 2006. "Re-thinking the Conceptualization of Customer Value and Service Quality within the Service-Profit Chain," *Managing Service Quality* (16:1), pp 23-36.

Watson, R.T., Pitt, L.F., and Kavan, C.B. 1998. "Measuring Information Systems Service Quality: Lessons from Two Longitudinal Case Studies," *MIS Quarterly* (22:1), pp 61-79.

## Acknowledgements

This work is supported by an Australian Research Council Linkage Project. We thank Mr. Paul Collins, CEO of Assessment Portal Pty Ltd for his input in providing an assessment platform for the SMPA approach. We also acknowledge the support we received from CITEC, a strategic ICT division for the Queensland Government and the Toowoomba Regional Council ICT department during the evaluation of the SMPA approach.

## Copyright